# Two-axis twisting using Floquet-engineered XYZ spin models with polar molecules

Calder Miller[1*], Annette N. Carroll[1], Junyu Lin[1], Henrik Hirzler[1], Haoyang Gao[2], Hengyun Zhou[2,3], Mikhail D. Lukin[2], Jun Ye[1*]

[1]JILA, National Institute of Standards and Technology, and Department of Physics, University of Colorado, Boulder, CO 80309, USA.

[2]Department of Physics, Harvard University, Cambridge, Massachusetts 02138, USA

[3]QuEra Computing Inc., 1284 Soldiers Field Road, Boston, MA, 02135, US

**Abstract**

Polar molecules confined in an optical lattice are a versatile platform to explore spin-motion dynamics based on strong, long-range dipolar interactions[1,2]. The precise tunability[3] of Ising and spin-exchange interactions with both microwave and dc electric fields makes the molecular system particularly suitable for engineering complex many-body dynamics[4–6]. Here, we used Floquet engineering[7] to realize interesting quantum many-body systems of polar molecules. Using a spin encoded in the two lowest rotational states of ultracold KRb molecules, we mutually validated XXZ spin models tuned by a Floquet microwave pulse sequence against those tuned by a dc electric field through observations of Ramsey contrast dynamics, setting the stage for the realization of Hamiltonians inaccessible with static fields. In particular, we observed two-axis twisting[8] mean-field dynamics, generated by a Floquet-engineered XYZ model using itinerant molecules in 2D layers. In the future, Floquet-engineered Hamiltonians could generate entangled states for molecule-based precision measurement[9] or could take advantage of the rich molecular structure for quantum simulation of multi-level systems[10,11].

**Introduction**

Periodic driving of a quantum system, known as Floquet engineering, can substantially modify the symmetries and dynamics of the system, leading to exciting new physics and applications. For example, dynamical decoupling[12,13], consisting of repeated spin-echo pulses, enables rejection of noise, extending coherence in quantum computing[14] and metrology[15] applications. By periodically modulating optical lattices, motion of atoms can be controlled[16], enabling simulation of gauge theories[17] and anyonic statistics[18]. By driving internal degrees of freedom in spin systems, time crystals[19,20] and tunable quantum magnets[21,22] have been created. Many other possible applications of Floquet engineering techniques are being explored, including the realization of topological insulators with ultracold molecules[23] and generation of metrologically useful GHZ-like[24] or spin-squeezed states[25].

Floquet engineering of interacting spin systems has been demonstrated in a range of physical platforms, including nuclear spins[26–28], superconducting qubits[29], Rydberg atoms[21,30], solid state defects[7,19,22,31], trapped ions[32], and ultracold molecules[5]. Optically trapped ultracold molecules[33,34] present a unique combination of features advantageous for studying many-body physics[1,2]. Scalable systems with tunable geometry and high-fidelity, state-resolved imaging of single particles[5,35–38] can be realized using optical tweezers[35,36,38–40] or lattices[41,42]. With appropriate trapping conditions[43–47], disorder and particle loss can be rendered much weaker than interactions, enabling study of highly coherent many-body systems. Rich rotational and hyperfine level structures allow tuning of interactions through choice of states[4] and simulation of higher-spin systems[48]. Microwave pulses enable fast, high-fidelity state control[4,41,49].

In prior work[4] using itinerant molecules, dc electric fields enabled tuning across a range of $U(1)$-symmetric XXZ interactions, ranging from pure spin exchange to an Ising-interaction dominated regime. However, less symmetric spin Hamiltonians exhibit interesting phases and dynamics, including efficient generation of spin-squeezed states through two-axis twisting (TAT)[8], but remain inaccessible with electric field tuning. Such systems can be explored through Floquet engineering, which can break $U(1)$ symmetry through rotation of the spins about arbitrary axes on the Bloch sphere.

Despite its broad utility, Floquet engineering has not been self-consistently validated against static Hamiltonians on an experimental platform. Such benchmarking could yield insight into how experimental imperfections limit Floquet engineering schemes. Here, we utilize the unique molecular level structure to validate that Floquet-engineered XXZ models produce similar Ramsey contrast decay dynamics, which probe energy shifts arising from interactions between molecules[41,50], to those tuned by electric field for both pinned and itinerant molecules. Our measurements reveal regimes where Floquet engineering works well, but also settings where its performance can be further improved. These verification protocols between different experimental implementations are important for the progress of quantum information science in general[51].

Building upon the comparison of XXZ spin dynamics, we designed and implemented a Floquet pulse sequence for generating TAT[8], which can produce spin-squeezed states with Heisenberg scaling applicable to precision measurements using molecules[9]. While TAT has been challenging to realize experimentally[52], our molecular system with strong dipolar interactions provides a surprisingly robust platform to implement this Hamiltonian via Floquet engineering. We characterized the dynamics of the Bloch vector under TAT using itinerant molecules, finding excellent agreement with a collective mean-field model.

**Floquet spin models**

The dynamics of a quantum system under periodic driving can be described with a Magnus expansion[53]. Its leading order for small $t_c/\tau$, where $t_c$ is the period of the cycle and $\tau$ is the characteristic timescale of interactions, is average Hamiltonian theory (AHT)[54]. Under AHT, the system evolves under the interaction picture Hamiltonian $H_{\text{avg}} = \frac{1}{t_c}\int_0^{t_c} H(t)dt$ where $H(t) = H(t - t_c)$ is the periodic Hamiltonian. If $H$ consists of periods of free evolution separated by rapid changes of the Hamiltonian, $H_{\text{avg}} = \frac{1}{t_c}\sum_i \tau_i H_i$, where $\tau_i$ is the time spent under Hamiltonian $H_i$ in each cycle. In spin systems studied experimentally, this can be realized by applying rapid microwave pulses that rotate each spin by an angle $\theta$ about an axis $\hat{n}$ on the Bloch sphere. Higher order corrections accounting for finite $\frac{t_c}{\tau}$ and pulse time can be accounted for in designing Floquet pulse sequences[15].

The so-called XYZ spin models of the form $H_{XYZ} = \sum_{i<j} J_{ij}(g_X s_i^X s_j^X + g_Y s_i^Y s_j^Y + g_Z s_i^Z s_j^Z)$, where $i$ and $j$ index the spins and $J_{ij}$ encodes the geometric scaling of the coupling strengths $\boldsymbol{g}$, describe fundamental phenomena in magnetism[3]. In particular, the subspace of XXZ models with $g_X = g_Y = g_\perp$, which can be parametrized by the anisotropy $\chi = g_Z - g_\perp$, includes the Ising ($g_\perp = 0$), XY ($g_Z = 0$), and Heisenberg ($\chi = 0$) models. Spins encoded in rotational levels $|0\rangle = |N = 0, m_N = 0\rangle$ and $|1\rangle = |1,0\rangle$ of polar molecules natively realize dipolar XXZ models with $J_{ij} \propto \frac{1-3\cos^2\theta_{ij}}{|r_{ij}|^3}$, where $\theta_{ij}$ is the angle between the quantization axis defined by the electric field $\boldsymbol{E}$ and the displacement $\boldsymbol{r}_{ij}$, $N$ is the rotational quantum number, and $m_N$ is its projection along the quantization axis[4,55]. At low electric fields, $g_Z \approx 0$, so the molecules predominantly interact via spin exchange $XX + YY$ interactions. The ratio of Ising interactions $g_Z$ to spin exchange interactions $g_\perp$ can be tuned by applying a dc electric field that mixes rotational states, inducing a dipole moment in the lab frame (Fig. 1A). Microwave pulses can rotate the state on the Bloch sphere, transforming the $XX + YY$ interaction into $XX + ZZ$ or $YY + ZZ$ interactions (Fig. 1B). By choosing the ratio of times $t_{X,Y,Z}$ spent in the different frames, a range of XYZ models can be realized (shaded triangle in Fig. 1C).

**Tunable XXZ Hamiltonians**

Dipolar XXZ spin models have been realized in a variety of platforms, including Rydberg atoms[21,30], magnetic atoms[56], nitrogen vacancy centers[22], and polar molecules. With molecules, bulk magnetization dynamics of the XY model in a disordered 3D lattice[41,50] were explored. Similar observations were recently made in a 2D system with a quantum gas microscope[5], where the evolution of spin correlations was observed under a Floquet pulse sequence that engineered an XXZ model at $|E| = 0$. Previously, we used electric fields to tune interactions between itinerant molecules in a stack of 2D layers[4], which at short times approximates an all-to-all spin model[57]. In Ref. 6, we studied electric field-tuned lattice spin models, including generalized $t$-$J$ models.

Here, we present a systematic study of a range of Floquet-engineered Hamiltonians, including an XYZ model for two-axis twisting. In doing so, we demonstrate the applicability of advanced pulse sequences for robust engineering of interactions and cancellation of disorder, explore the efficacy of Floquet engineering in systems involving both spin dynamics and motion, and cross verify the Floquet engineering and electric field tuning of Hamiltonians, showing the broad utility of this approach. We first explore XXZ spin models in a 3D lattice, which can be compared to measurements in Ref. 6 using electric field tuning. Our system is again characterized using Ramsey spectroscopy. Molecules were produced in the rovibrational ground state $|0\rangle$. A microwave $\pi/2$ pulse on the $|0\rangle \leftrightarrow |1\rangle$ transition prepared them in the state $|+Y\rangle^{\otimes N}$, after which they evolved for a time $t$, during which a microwave pulse sequence was applied, before another $\pi/2$ pulse with phase $\phi$ completed the Ramsey sequence. We then measured the numbers of molecules in $|0\rangle$ and $|1\rangle$[42]. The fraction of the molecules in $|1\rangle$, $f_1$, was measured twice for 8 values of $\phi$ equally spaced between 0 and $2\pi$ radians, then the Ramsey contrast was computed from the standard deviation of $f_1$ as in Ref. 42. This process allows the length of the Bloch vector to be measured even when the phase is randomized shot to shot at longer evolution times.

To generate XXZ Hamiltonians with Floquet engineering, the molecules were produced at $|E| = 1$ kV/cm, where the interaction Hamiltonian is still predominantly spin-exchange, but where the hyperfine and rotational structure is sufficiently split that numerous pulses on the $|0\rangle \leftrightarrow |1\rangle$ transition can be driven with only modest loss to other states. Instead of the Knill dynamical decoupling (KDD) sequence[14] used in Ref. 6, which only removes single-particle dephasing, we used a DROID-R2D2 pulse sequence[58], which also can tune XYZ interactions, during the evolution time. We used an average pulse spacing of 100 us, which optimizes the coherence time for an XXX model (see Methods). By varying the time in the different frames in the sequence, the interaction Hamiltonian can be tuned between approximately $g_{\perp,z} = (g_\perp^0, 0)$ and $g_{\perp,z} = \left(\frac{g_\perp^0}{2}, g_\perp^0\right)$ (dashed blue trace in Fig. 2A), which corresponds to $|E| = 10.5$ kV/cm. Here, $g_\perp^0$ is the native spin-exchange interaction. For comparison, in Ref. 6, by varying $|E|$ between 1 and 12.7 kV/cm, the coefficients $g_{\perp,z}$ of the XXZ Hamiltonian were tuned between $2\pi \times (210, 6)$ Hz and $2\pi \times (75, 177)$ Hz for molecules on adjacent lattice sites in the plane perpendicular to $E$ (solid red trace in Fig. 2A).

The contrast was then measured for different $t$ by varying the number of times the pulse sequence was repeated. The contrast decay curve is fit with a stretched exponential model $C(t) = e^{-(\Gamma t)^\nu}$, where $\Gamma$ is the contrast decay rate and $\nu$ is the stretching parameter describing sub-exponential decay due to number loss or glassy dynamics[6,59]. Sample contrast decay traces for initial densities of approximately $1.6 \times 10^7$ cm$^{-2}$ are shown in Fig. 2B for spin exchange interactions (blue circles), Floquet-engineered XXX interactions (orange squares) and XXX interactions realized with electric field (green triangles). The contrast decay is much slower for both realizations of XXX interactions than for the spin-exchange interactions. The contrast decay for the Floquet-engineered XXX interaction is slightly faster than for the electric field-tuned XXX interaction, possibly because of a slower rate of dynamical decoupling pulses or higher order terms not fully symmetrized by the Floquet engineering.

To ensure that the measured contrast decay is due to dipolar many-body interactions and not particle loss or one-body dephasing, we measured contrast decay curves at several initial average densities $n$ (see Methods). In Fig. 2C, the fitted $\Gamma$ is plotted as a function of 2D density for spin exchange interactions (blue circles) and Floquet-engineered XXX interactions (orange squares). We fit a linear model $\Gamma = \kappa n + \Gamma_0$ to the data (lines in Fig. 2C), extracting the slope $\kappa$, which describes the density dependence of contrast decay due to interactions. Linear fits to the data show a much stronger density dependence of the contrast decay rate for the spin-exchange model than the XXX model, which has a vanishing density dependence within measurement uncertainty. The nonzero y-intercepts $\Gamma_0$ are a measure of residual single particle dephasing.

We equate the XXZ Hamiltonians realized by electric fields and Floquet engineering by determining the value of $|E|$ that yields the same ratio of $g_\perp$ to $g_Z$ as a set of Floquet timings (Fig. 2A). Because the overall strength of the interaction decreases for higher electric fields but remains constant for the Floquet sequences, we rescale the Floquet data's $\kappa$ by the ratio of $g_\perp$ under $|E|$ to its value under the Floquet Hamiltonians. We plot $\kappa$ as a function of $\chi$ for both datasets (Fig. 2D). There is good agreement between their dynamics, suggesting that Floquet engineering is realizing the desired spin models. When the molecules are confined in a deep 3D lattice, we find that $\kappa$ changes approximately linearly as a function of $|\chi|$, decreasing to approximately zero at the Heisenberg point where $\chi = 0$, then increasing as the Ising interactions come to dominate. This trend agrees with a moving average cluster expansion simulation[50] of the contrast decay from Ref. 6.

To investigate the limitations of Floquet engineering, we applied the same pulse sequence to itinerant molecules confined in a 1D optical lattice (Fig. 2E), where motion and collisions complicate the Hamiltonian[6]. In the spin exchange regime where the Floquet engineering only slightly modifies the Hamiltonian, we observe excellent agreement between the electric field-tuned and Floquet-engineered dynamics. However, in the Heisenberg and Ising regimes, the Floquet-engineered model shows a higher rate of density-dependent contrast decay than the electric field-controlled version. We expect this to occur since the contrast decay in an itinerant system is dominated by short-range collisions between molecules[4], which was modeled in Ref. 6 by a Monte Carlo simulation (solid line in Fig. 2E) incorporating two-particle scattering. The collisions take place over durations comparable to the Floquet pulse spacing and have scattering properties set by the electric field. Nevertheless, a similar overall trend is observed with Floquet engineering, indicating that control is still possible over itinerant models.

**Realizing two-axis twisting**

Beyond providing an alternative to electric field tuning to realize XXZ models, Floquet pulse sequences can engineer general XYZ models, which breaks the symmetries of models realized using only a static electric field. One special case, with $g_X = -g_Z$, $g_Y = 0$, generates TAT dynamics. TAT can achieve spin squeezing with number scaling reaching the Heisenberg limit, exceeding the performance of one-axis twisting (OAT) generated by XXZ Hamiltonians[8,60]. Generation of spin-squeezed states with ultracold molecules could enhance applications in precision metrology, including detection of beyond-standard-model physics[9]. However, despite several proposals[25,60], TAT mean-field dynamics have only recently been observed in a cavity QED system[52], since the required XYZ Hamiltonians are not native to common platforms.

Ultracold molecules in a 1D optical lattice provide a platform with sufficient control to study TAT. We implemented TAT and observed the evolution of Bloch vectors under the Hamiltonian by Floquet engineering the interactions between itinerant molecules in two-dimensional layers, which approximate all-to-all interactions[4,57]. The native spin-exchange interactions generate OAT about the $Z$ axis (Fig. 3C) plus an isotropic XXX interaction, with Hamiltonian $H_{\text{OAT}} = g_\perp \sum_{i<j} J_{ij}(\mathbf{s}_i \cdot \mathbf{s}_j - s_i^Z s_j^Z)$. The Hamiltonian is invariant under dynamical decoupling sequences consisting solely of $\pi$ pulses such as XY8 (Extended Data Fig. 1A). Because rotations on the Bloch sphere preserve the trace $g_X + g_Y + g_Z$ of the interaction Hamiltonian, the native low field Hamiltonian $g_\perp > 0, g_Z \approx 0$ cannot be transformed into pure TAT. Instead, we realize the Hamiltonian $H_{\text{TAT}} = g_\perp \sum_{i<j} J_{ij} \left(\frac{2}{3}\mathbf{s}_i \cdot \mathbf{s}_j + \frac{1}{3}\left(s_i^X s_j^X - s_i^Z s_j^Z\right)\right)$, as proposed in Ref. 25. Since any point on the surface of the Bloch sphere is an eigenstate of the $\vec{s}_i \cdot \vec{s}_j$ term in $H_{\text{OAT}}$ and $H_{\text{TAT}}$, the term does not affect the short-time dynamics. To engineer TAT while cancelling inhomogeneous disorder, we used a modified XY8 pulse sequence, XY8-TAT (Extended Data Fig. 1B), in which the $\pi$ pulses about the Y axis are split into pairs of $\pi/2$ pulses. For pulse spacing $\tau$, this results in $4\tau$ being spent with $XX + ZZ$ interactions and $8\tau$ with $XX + YY$ interactions, realizing $H_{\text{TAT}}$ on average (see Methods).

We first prepared the spins in $|0\rangle$ and applied a pulse with area $\theta$ about axis $-\sin\phi \hat{X} + \cos\phi \hat{Y}$ on the Bloch sphere, giving an initial average Bloch vector $\langle \mathbf{S}_0 \rangle = \frac{1}{2}(\cos\phi \sin\theta, \sin\phi \sin\theta, \cos\theta)$. We then repeated either XY8-TAT or a standard XY8 pulse sequence for 2.4 ms before measuring the collective Bloch vector in the $X, Y$ or $Z$ bases. This measurement requires excellent shot-to-shot phase coherence, so we operated at $|\mathbf{E}| = 0$ to reduce fluctuations in the transition frequency from electric field noise. The data is compared to a mean-field, all-to-all simulation of the dynamics (see Methods) where the interaction strength and dephasing rate are fit to the data.

One-axis twisting has two stable fixed points at the $\pm\hat{Z}$ poles of the Bloch sphere. Between the poles, interactions rotate the Bloch vector about the $Z$ axis (Fig. 3A). As observed in previous work [4], the phase shift $\Delta\phi = \arctan\frac{\langle S^Y \rangle}{\langle S^X \rangle} - \arctan\frac{\langle S_0^Y \rangle}{\langle S_0^X \rangle}$ under OAT is proportional to $\langle S_0^Z \rangle$ (Fig. 3C). We also see that $\langle S^Z \rangle$ does not depend on $\phi$ (Fig 3E). By contrast, TAT has two unstable fixed points at the $\pm\hat{Y}$ poles of the Bloch sphere and four stable fixed points where the $\pm X$ and $\pm Z$ axes intersect the $Y = 0$ plane[61,62]. Between the fixed points, the Bloch vector moves alternately upward and downward as a function of angle in the $X - Z$ plane, completing two oscillations around the circumference of the sphere (Fig. 3B). This behavior is apparent in our data (Fig. 3F).

Two-axis twisting can prepare spin squeezed states from initial coherent states at the poles of the Bloch sphere by compressing and extending the quasiprobability distribution in orthogonal directions rotated by $\pi/4$ radians relative to the twisting axes[61] (Fig. 4A). We observed the mean-field version of these dynamics by preparing a ring of initial states in an X-Z plane of the Bloch sphere for different values of $\langle S_0^Y \rangle$ and fitting their positions after evolution under TAT for 2.4 ms to an ellipse (positions before (after) evolution in blue (orange) are shown in insets in Fig. 4B). We plot the ratio of the sizes of the fitted ellipses along the $X = \pm Z$ directions as a function of $\langle S_0^Y \rangle$ (Fig. 4B). Near $\langle S_0^Y \rangle = \pm 0.5$, the ellipses are more stretched, with the major axis in opposite directions. The measured shapes agree well with a mean-field simulation of the experiment (solid line in Fig. 4D). The reduced size of the ellipse after time evolution relative to the initial circle is likely a result of dephasing from collisions and inhomogeneous dynamics between the layers.

Near the $\pm Y$ poles, the rate of rotation of the Bloch vector is proportional to the Bloch vector's angle from the pole, resulting in exponential growth of the displacement with time. We measured the change in the angle $\theta_Y$ between $\langle \mathbf{S} \rangle$ and the $\hat{Y}$ axis for an evolution time of 2.4 ms after preparing states at a range of $\langle S_0^Y \rangle$ values for $\langle S_0^X \rangle = \pm\langle S_0^Z \rangle$, where $\frac{d\theta_Y}{dt}$ should be extremized, and $\langle S_0^X \rangle = 0$ where it should be zero (green, purple, and gray lines in Fig. 4C, respectively). We observe good agreement with the mean-field model at 2.4 ms evolution time (green ($\langle S_0^X \rangle = \langle S_0^Z \rangle$), purple ($\langle S_0^X \rangle = -\langle S_0^Z \rangle$), and gray ($\langle S_0^X \rangle = 0$) points and solid lines in Fig. 4D). Because of the finite time, $\Delta\theta_Y$ (gray points in Fig. 4D) is nonzero for $\langle S_0^X \rangle = 0$, although the $\frac{d\theta_Y}{dt} = 0$ at $t = 0$.

**Outlook**

In this work, we systematically characterize Floquet engineering methods for controlling interactions between ultracold molecules and demonstrate their application to realizing spin Hamiltonians applicable to metrology and many-body physics. Comparing Ramsey contrast dynamics produced by Floquet engineering against those generated by a dc electric field helps verify its efficacy. These Hamiltonian engineering tools will enable diverse future research. If an interacting ensemble was prepared in a single layer of an optical lattice and detection efficiency was improved, it should be possible to verify entanglement of a spin-squeezed state prepared via OAT or TAT through noise measurements[63]. Time-reversal of the dynamics[64,65], implemented using another rotational level[4], may enable quantum-enhanced sensing with relaxed detection requirements. More generally, in combination with tunneling in an optical lattice and evaporative cooling[66], Floquet engineering can realize highly tunable *t-J* and Hubbard models, allowing investigation of phenomena in superfluidity, quantum magnetism[3,6,55], or topological matter[23]. The rich level structure of molecules will also enable investigation of dynamics of higher spins, for which similar Hamiltonian engineering techniques[11] exist, and exploration of synthetic dimensions encoded in rotational states[10].

**Figures**

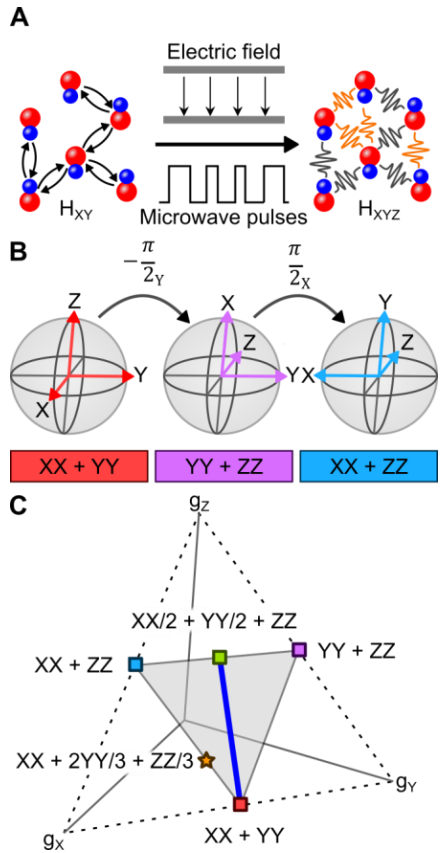

**Fig. 1. KRb Hamiltonian engineering.** (**A**) At low electric fields, molecules interact via spin exchange. XXZ and XYZ models can be realized with dc electric fields and microwave pulse sequences, respectively. (**B**) Microwave $\pi/2$ pulses rotate the spins on the interaction picture Bloch sphere, transforming the spin exchange XX + YY interactions into XX + ZZ or YY + ZZ interactions. (**C**) A range of XXZ (blue line) and XYZ (shaded triangle) interaction Hamiltonians can be realized by spending different amounts of time in the different frames, starting with the low-electric-field spin exchange Hamiltonian between the $|0\rangle$ and $|1\rangle$ states. The orange star indicates the Hamiltonian used to study two-axis twisting dynamics. The axes represent the strength of the couplings in the XYZ Hamiltonian.

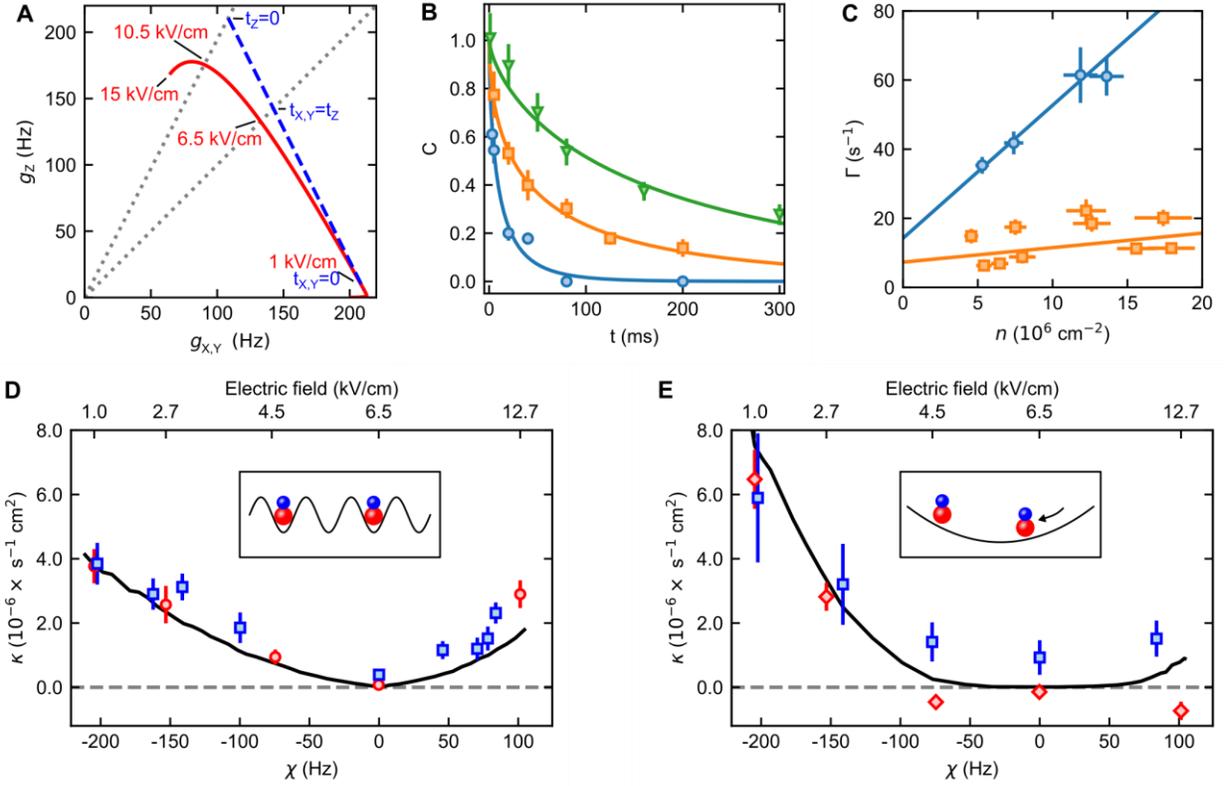

**Fig. 2. Benchmarking XXZ spin dynamics.** (**A**) XXZ Hamiltonians can be prepared with electric fields (red solid line) and Floquet engineering (blue dashed line). The dashed diagonal lines show the subspaces of coefficients with equal ratio $g_\perp/g_Z$ corresponding to the most Ising Hamiltonian that can be prepared with Floquet engineering (left) and the Heisenberg XXX model (right). (**B**) Ramsey contrast as a function of time is shown for a spin exchange model (blue circles), a Floquet engineered XXX model (orange squares) and an XXX model realized with electric fields (green triangles) with initial densities of about $1.6 \times 10^7$ cm$^{-2}$. The lines are fits to a stretched exponential. The error bars on the data are 1 s.d. from bootstrapping. (**C**) The fitted contrast decay rate is plotted as a function of 2D density for a spin exchange model (blue circles) and a Floquet engineered XXX model (orange squares). The $\Gamma$ error bars are 1 s.e. from stretched exponential fits and the $n$ error bars are 1 s.e. from the initial density of an exponential fit. (**D**) The density-normalized contrast decay rate $\kappa$ is plotted as a function of $\chi = g_Z - g_\perp$ and (effective, in the case of the Floquet data) electric field for simulations (black line), electric field data (red circles) and Floquet data (blue squares) for molecules pinned in a deep 3D lattice (schematic in inset). The error bars are 1 s.e. from a linear fit. (**E**) Same as (D), but for itinerant molecules in a 1D vertical lattice. The electric field-tuned data and simulations in this figure are from Ref. 6.

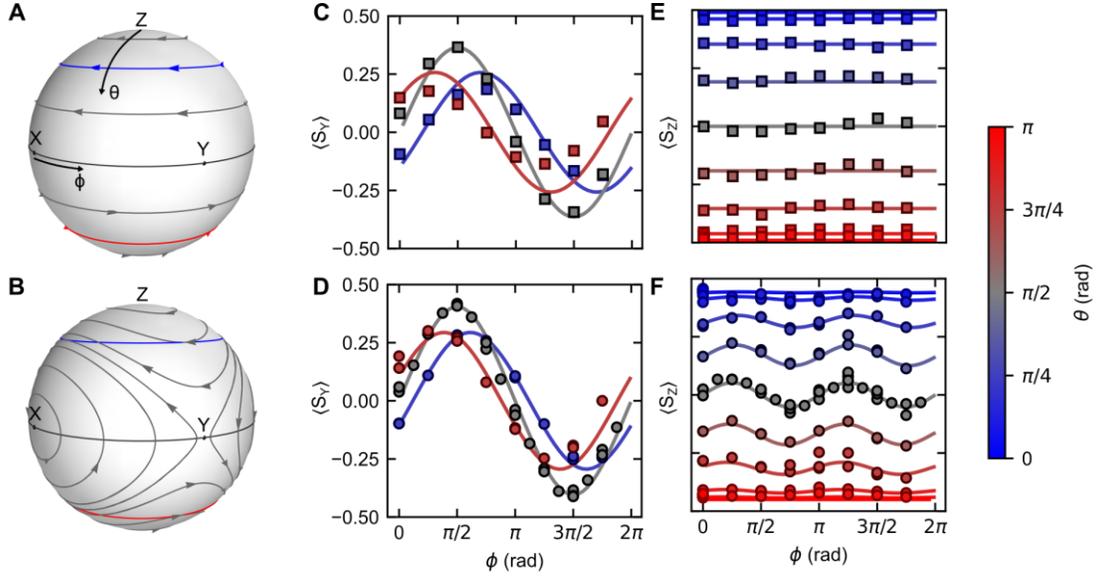

**Fig. 3. Engineering one-axis and two-axis twisting.** The Bloch vector phase portraits for OAT (**A**) and TAT (**B**) are shown on the Bloch sphere. The blue, dark gray, and red latitudinal lines mark $\theta = \frac{\pi}{4}, \frac{\pi}{2}, \frac{3\pi}{4}$ respectively. The projection of the average Bloch vector in the $Y$ (**C** and **D**) and $Z$ (**E** and **F**) directions is shown for OAT (**C** and **E**) and TAT (**D** and **F**) as a function of the phase (x axis) and tipping angle (color) of the initial state after 2.4 ms evolution. The points are experimental data, and the solid lines are results from a mean-field simulation.

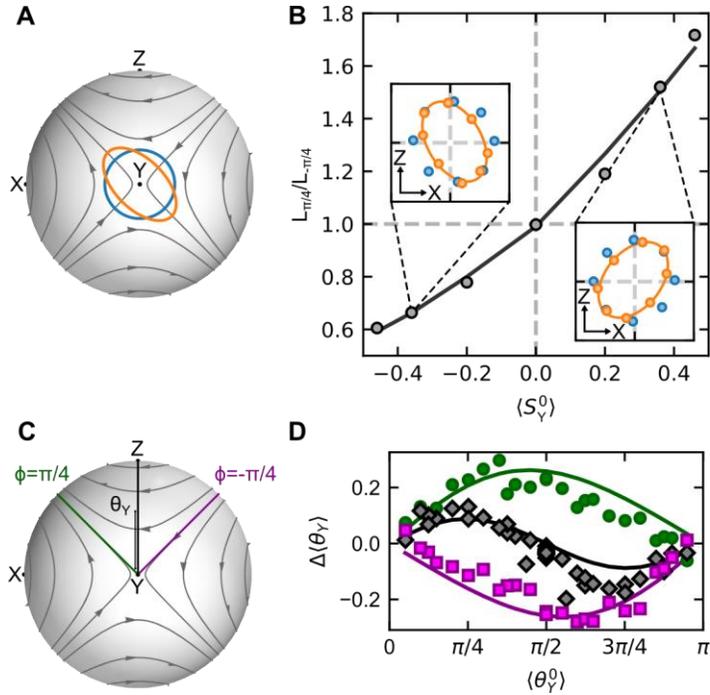

**Fig. 4. Two-axis twisting mean-field dynamics.** (**A**) TAT mean-field dynamics are plotted around the +Y pole of the Bloch sphere, where spin squeezing would be generated. In a mean-field analog to squeezing, points along the blue circle evolve to points along the orange ellipse.

Similar dynamics mirrored about the Y-Z plane would be observed near the -Y pole. (**B**) An ellipse is fit to data (orange points in insets) prepared at different phases $\phi$ in the X-Z plane and angles $\theta_Y^0$ from the +Y axis (blue points in inset) after 2.4 ms evolution time. The aspect ratio of the ellipse, as measured along axes with angle $\frac{\pi}{4}$ and $-\frac{\pi}{4}$ from the +Z axis, is plotted as a function of the initial spin $\langle S_Y^0 \rangle = \frac{\cos \theta_Y^0}{2}$ (black points). The same fit to the mean-field simulations is plotted as the black line. The insets show the X and Z components of the initial and final Bloch vectors for $\theta_Y^0 = \pi/4$ and $3\pi/4$. The axis range is -0.5 to 0.5. (**C**) TAT mean-field dynamics are plotted around the +Y pole of the Bloch sphere. The contours at $\phi = -\frac{\pi}{4}, 0$ and $\frac{\pi}{4}$ are plotted in green, dark gray, and purple, respectively. (**D**) Measured values of $\langle \theta_Y \rangle$ after 2.4 ms evolution under TAT dynamics is plotted as a function of $\theta_Y^0$ for $\phi = \frac{\pi}{4}$ (green circles), 0 (gray diamonds), and $-\frac{\pi}{4}$ (purple squares). The lines are mean-field simulation results under the same conditions.

## Methods

### Molecule production

We produced fermionic $^{40}$K$^{87}$Rb (KRb) molecules from quantum degenerate gases of potassium and rubidium by magnetoassociation followed by stimulated Raman adiabatic passage (STIRAP) to the rovibrational ground state[34,67,68]. The molecules were produced in a 3D lattice formed from 1064 nm light with a spacing of $a_{x,y,z} = (532, 540, 532)$ nm, where $y$ is the direction of gravity and our dc electric and magnetic fields. Due to harmonic confinement from the lattice and an additional crossed dipole trap, the positions of the molecules approximated a Gaussian distribution with a standard deviation of $\sigma_{x,y,z} \approx (16, 2.8, 20)$ microns. The vertical extent of the distribution was measured to be $L = \frac{N^2}{\sum_k N_k^2} = 18.5(10)$ layers with layer-resolved gradient spectroscopy[42]. Here, $N$ is the total molecule number and $N_k$ is the number of molecules in the $k$th layer.

For the XXZ contrast decay measurements, around 12000 molecules were created in a lattice of depth $U_{x,y,z} = (25, 65, 25) E_r$, yielding a maximum average filling fraction of about 13%. Following STIRAP, the lattice was ramped to $U_{x,y,z} = (65, 65, 65) E_r$, where $E_r$ is the recoil energy, in 5 ms to suppress tunneling in the x-z plane for the pinned measurements or $U_{x,y,z} = (0, 65, 0)$ for the itinerant measurements. For the OAT and TAT measurements, we used a vertical lattice of depth 40 $E_r$, the shallower vertical lattice resulting in lower temperature and reduced ac Stark shifts from the lattice's non-magic polarization[42,43]. About 10000 molecules were produced in 2D layers at a temperature of 158(10) nK. For all measurements, the molecules were made and imaged at the target electric field, with the frequencies of the STIRAP lasers tuned to remain resonant with the molecular transitions.

To prepare systems at different densities, we used a microwave pulse with area $\theta$ to shelve partially the molecules in $|1\rangle$ while removing the molecules in $|0\rangle$ using a pulse of resonant light from the STIRAP down leg[4]. This procedure reduced the density by a factor of $\sin^2 \theta/2$.

## Microwave control of rotational states

We encoded a spin 1/2 system in the rotational states $|0\rangle = |N = 0, m_N = 0, m_K = -4, m_{Rb} = 1/2\rangle$ and $|1\rangle = |N = 1, m_n = 0, m_K = -4, m_{Rb} = 1/2\rangle$. The frequency of the $|0\rangle \leftrightarrow |1\rangle$ transition varies between 2.228 GHz at zero electric field and 4.225 GHz at $|\boldsymbol{E}| = 12.7$ kV/cm. We generated microwaves to drive Rabi oscillations between $|0\rangle$ and $|1\rangle$ using a custom FPGA-based RF synthesizer whose output was mixed with a microwave local oscillator. The microwaves were bandpass filtered, amplified, and then coupled to the molecules through the in-vacuum electrodes using a bias tee. We used rectangular pulses with Rabi frequency $\Omega \approx 2\pi \times 100$ kHz except at $|\boldsymbol{E}| = 0$ kV/cm, where we reduced $\Omega$ to $2\pi \times 50$ kHz to suppress off-resonant driving to other rotational and hyperfine states.

To reduce inhomogeneous broadening, we used magic angle optical traps for the crossed dipole trap and the two horizontal lattices[42,43]. Due to geometric constraints, the vertical lattice was not magic.

We characterized the fidelity of our microwave pulses using the one qubit randomized benchmarking sequence described in Ref. 69. At 1 kV/cm, we measured fidelities of 0.99941(9) per $\pi/2$ pulse with itinerant molecules confined by a 65 $E_r$ vertical lattice and 0.99915(14) with molecules pinned in a 65 $E_r$ 3D lattice. At 0 kV/cm, we measured a lower fidelity of 0.9941(9) with itinerant molecules, which we attribute to off-resonant driving of other nearby hyperfine states.

A substantial fraction of the molecules are lost during the application of hundreds or thousands of microwave pulses, particularly at 0 kV/cm. Because we image both $|0\rangle$ and $|1\rangle$, we can reject the effect of these erasure errors[70]. During our TAT pulse sequence, the number decays with an exponential time constant of 11.2(6) ms (Extended Data Fig. 4), which is long compared to the 2.4 ms measurement time.

## Optimizing Floquet timing

To optimize the pulse spacing for the Floquet XXZ data, we measured the contrast decay rate for an XXX Hamiltonian with pinned molecules, realized with a variety of pulse spacings between 25 and 400 us (Extended Data Fig. 3). If the pulse spacing is too long, interactions between the molecules will not be well symmetrized and dephasing from inhomogeneous single-particle noise will not be effectively canceled. If the pulse spacing is too short, pulse errors can accumulate more quickly and more molecules will be driven into other hyperfine states, causing increased loss. We therefore used a pulse spacing of 100 $\mu$s for the Floquet XXZ and XYZ Hamiltonians, which was the longest spacing that shows near-optimal contrast decay rates. For the KDD pulse sequence in Ref. 6, we used a 50 $\mu$s pulse spacing, as the molecules are more sensitive to fluctuating electric fields as the dipole moment increases at larger fields, so faster dynamical decoupling improves contrast.

## Imaging and image analysis

We imaged both the $|0\rangle$ and $|1\rangle$ states of the molecules using state selective STIRAP[4,42]. At all electric fields used except 1 kV/cm, we transferred the population in $|0\rangle$ to the Feshbach state, then detected the potassium atoms in the Feshbach molecules by absorption imaging. The reported number derives from a 2D Gaussian fit to the in-situ distribution of the atoms. We then applied a $\pi$ pulse to transfer the molecules in $|1\rangle$ to $|0\rangle$ and repeated the STIRAP and absorption imaging. At 1 kV/cm, the STIRAP depletes the molecules in $|1\rangle$, so we shelved the population in $|1\rangle$ to $|N=2, m_N=0\rangle$ during the first STIRAP. To calibrate the STIRAP efficiency, we performed repeated STIRAP transfers between the Feshbach state and $|0\rangle$ and fit an exponential decay to the molecule number. The one-way STIRAP efficiency is approximately 90% at 0 kV/cm, 85% at 1 kV/cm, and as low as 70% at 12.7 kV/cm and the efficiency of imaging Feshbach molecules is approximately 70%[68]. Both efficiencies are factored into the reported molecule number and densities. The reported 2D densities are computed assuming the average molecule number in a layer $N_{2D} = N/L$ and a Gaussian distribution with the fitted size from in-situ images.

We also characterized systematic errors introduced during the imaging process by preparing a superposition $\cos\theta\,|0\rangle + \sin\theta\,|1\rangle$ with a microwave pulse immediately before our imaging sequence. For itinerant molecules, we find that for $\theta$ close to 0 or $\pi$ fewer molecules are in the minority state than would be expected from the state preparation. We attribute this to loss from inelastic collisions between the Ramsey sequence and imaging. Since these collisions occur nearly 100 times more rapidly for distinguishable molecules[42], their effect on the density can be modeled as $\frac{dn_i}{dt} \propto -\beta_p n_i n_j$ for $i \neq j$ where the indices are over the states $|0\rangle$ and $|1\rangle$. By fitting the solution to this system of differential equations to the measured molecule numbers, we correct the density and spin data for this loss process (Extended Data Fig. 2).

**Mean-field OAT and TAT models**

Per the Ehrenfest theorem[71], the expectation values of the collective Bloch vector $\mathbf{S}$ under the all-to-all Hamiltonian $H_{XYZ} = \sum_i g_i S_i^2$ evolve as $\frac{d\langle S_i\rangle}{dt} = 2\sum_{j,k} \epsilon_{ijk}\, g_j \langle S_j\rangle\langle S_k\rangle$ where $i, j, k \in \{\hat{X}, \hat{Y}, \hat{Z}\}$. We also include dephasing with rate $\gamma^\perp$, adding a term $-\gamma^\perp \langle S_i\rangle$ to the equations for $\frac{d\langle S_X\rangle}{dt}$ and $\frac{d\langle S_Y\rangle}{dt}$. We numerically solve the equations of motion to obtain final values for $\langle S_i\rangle$.

For OAT, $g_{X,Y,Z} = (g, g, 0)$ and for TAT, $g_{X,Y,Z} = (3g, 2g, g)/3$ under AHT. When $\gamma^\perp$ is set to zero, we observe excellent agreement between simulations using the AHT interaction strengths and simulations of applying the XY8-TAT pulse sequence with XY interactions. To properly simulate the dephasing, which is partially transformed into depolarization by the Floquet pulse sequence, we chose to model the TAT dynamics by modeling the application of the XY8-TAT pulse sequence with instantaneous pulses and XY interactions between pulses.

To find the values of $g$ and $\gamma$, we fit the output of our simulations to the OAT and TAT data. To ensure that the fit is not dominated by points with abnormal molecule number, we only included measurements where the number is within 2 standard deviations of the average number for the TAT data. We found that the sum of squares of the errors was minimized when $g = 360$ s$^{-1}$ and $\gamma^\perp = 130$ s$^{-1}$ for our typical molecule density of $1.08(9) \times 10^7$ cm$^{-2}$.

**Methods References:**

**Acknowledgments:**

We thank Nathaniel Leitao, Leigh S. Martin, and Ana Maria Rey for helpful discussions and Krzysztof P. Zamarski and Felix Vietmeyer for technical contributions. We thank Kyungtae Kim and Chengyi Luo for their comments on the manuscript.

The work at JILA and Harvard was jointly supported by the US Department of Energy, Office of Science, National Quantum Information Science Research Centers, Quantum Systems Accelerator. Support is also acknowledged from NSF QLCI OMA-2016244, NSF PFC PHY-2317149, AFOSR MURI, ARO MURI, and NIST. H.G., H.Z., and M.D.L. acknowledge support from the Center for Ultracold Atoms, an NSF Physics Frontiers Center. C.M. acknowledges support from the Department of Defense through the National Defense Science and Engineering Graduate Fellowship. A.N.C. acknowledges support from the National Science Foundation Graduate Research Fellowship under Grant No. DGE 2040434.

**Author contributions:**

C.M., A.N.C., J.L., H.H., and J.Y. performed the experiments and analyzed the data. H.G., H.Z., and M.D.L. contributed ideas and methods for Floquet engineering. All authors discussed the results and contributed to the manuscript.

**Competing interests:** The authors declare no competing interests.

**Data availability:** Data are available from the corresponding authors upon request.

**Code availability:** Code is available from the corresponding authors upon request.

**Additional information:**

Correspondence and requests for materials should be addressed to Calder Miller (calder.miller@colorado.edu) and Jun Ye (ye@jila.colorado.edu).

## Extended Data Figures

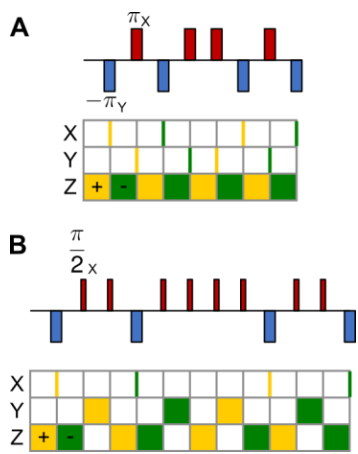

**Extended Data Fig. 1. Pulse sequences** Pulse sequences for generating OAT (XY8; **A**) and TAT (XY8-TAT; **B**) dynamics are shown using the notation of Ref. 7. Narrow (wide) rectangles represent $\pi/2$ ($\pi$) pulses, red (blue) rectangles pulses about the ±X (±Y) axes, and pulses above (below) the line about the + (-) axes. The frame matrix representation shows which axis points along the +Z direction as a function of time, with yellow (green) blocks representing axes originally along the + (-) directions.

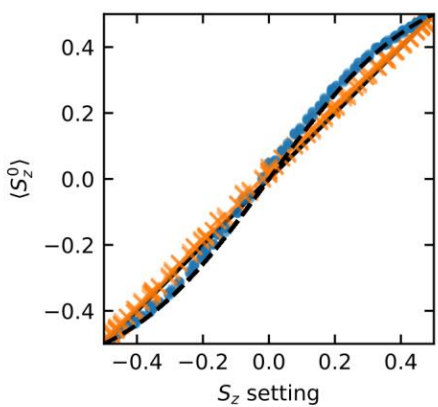

**Extended Data Fig. 2. Imaging correction** The measured value of $\langle S_z^0 \rangle$ is plotted (blue circles) as a function of the value prepared by a microwave pulse. The dashed curve is a fit to the loss model (see Methods). By inverting the model, the data can be corrected for loss during imaging (orange crosses).

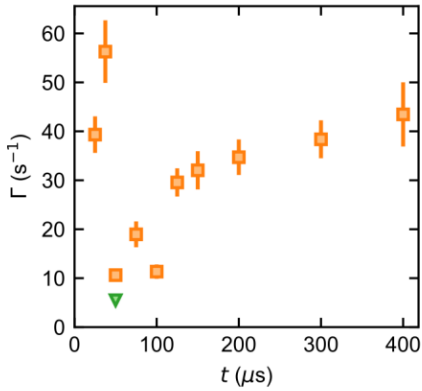

**Extended Data Fig. 3. Optimizing Floquet pulse timing** The fitted contrast decay rate $\Gamma$ is plotted as a function of pulse spacing $t$ for molecules with initial densities around $1.8(5) \times 10^7$ cm$^{-2}$ in a 3D lattice with parameters set to produce the XXX model. The green point is the electric field-tuned data with a KDD pulse sequence and the orange points are the Floquet data with a DROID-R2D2 pulse sequence. The error bars on the plot are 1 s.e. from stretched exponential fits.

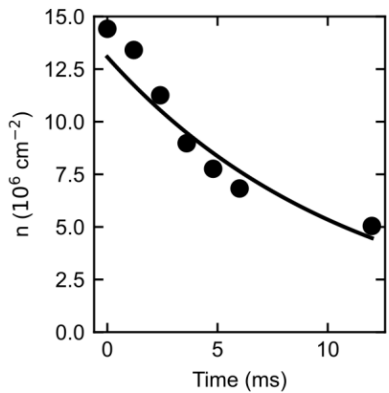

**Extended Data Fig. 4. Number loss during TAT Floquet engineering** The average molecule density is plotted as a function of time as the XY8-TAT pulse sequence is repeatedly applied. The solid curve shows an exponential fit to the data, with time constant 11.2(6) ms.

20